\documentclass[%
 aip,
 amsmath,amssymb,
reprint,%
author-numerical,%
]{revtex4-1}

\usepackage{graphicx}
\usepackage{dcolumn}
\usepackage{bm}

\usepackage[utf8]{inputenc}
\usepackage[T1]{fontenc}
\usepackage{mathptmx}
\usepackage{etoolbox}

\makeatletter
\def\@email#1#2{%
 \endgroup
 \patchcmd{\titleblock@produce}
  {\frontmatter@RRAPformat}
  {\frontmatter@RRAPformat{\produce@RRAP{*#1\href{mailto:#2}{#2}}}\frontmatter@RRAPformat}
  {}{}
}%
\makeatother
\begin{document}

\preprint{AIP/123-QED}

\title{Low-coherence interferometric measurement of the spectral dependence of the light field backscattered by optical interfaces}
\author{M. Lequime}
\author{I. Khan}%
\author{A. Bolliand}%
\author{M. Zerrad}%
\email{myriam.zerrad@fresnel.fr}
\author{C. Amra}%
\affiliation{ 
Aix Marseille Univ, CNRS, Centrale Marseille, Institut Fresnel, Marseille, France
}%

\date{\today}

\begin{abstract}
In this paper, we show how the combined use of low coherence interferometry, balanced detection and data processing comparable to that used in Fourier transform spectrometry allows us to characterize with ultimate resolutions (sub-ppm in level, 0.2 nm in wavelength and 5 mdeg in angle) the retro-reflection and retro-scattering response of both sides of a 2 mm thick silica wafer.
\end{abstract}

\maketitle


Scattered light is one of the most important disturbance factors that can limit the ultimate sensitivity of giant laser interferometers developed for gravitational wave detection\cite{Canuel_2013,Spector_2012}. This is especially true when this scattered light is superimposed on the main beam with a time-varying phase term, as the resulting spurious signal, if falls within the operating frequency band of the interferometer, can degrade the detection quality of a signal generated by the passage of a gravitational wave\cite{Ottaway_2012}.

It is therefore very important to be able not only to estimate theoretically the intensity and angular distribution of this scattered light inside the interferometer\cite{Vinet_1997,Vander-Hyde_2015,Zeidler_2017}, but also to measure experimentally the fraction of this scattered light that couples coherently to the main beam, first at the level of elementary components\cite{Magana-Sandoval_2012, Was_2022}, and then within the instrument itself\cite{Was_2021}. In this paper, we describe an interferometric method that allows the recording of such a measurement not only from elementary components but also from more complex subsystems.

This paper is organized as follows: after describing the experimental setup used for the acquisition of the data related to this measurement, we detail the digital processing implemented to improve the signal-to-noise ratio of these raw data and to determine the spectral dependence of the amplitude of the field retro-propagated by the sample. We then present the result of these measurements when the sample is a silica wafer with, on one side, a V-shaped antireflection coating centered at 1055 nm. In conclusion, we make a critical analysis of these first results and consider possible ways to improve this new measurement method and to extend its use to more complex systems.

\begin{figure*}
\includegraphics[width=1.00\textwidth]{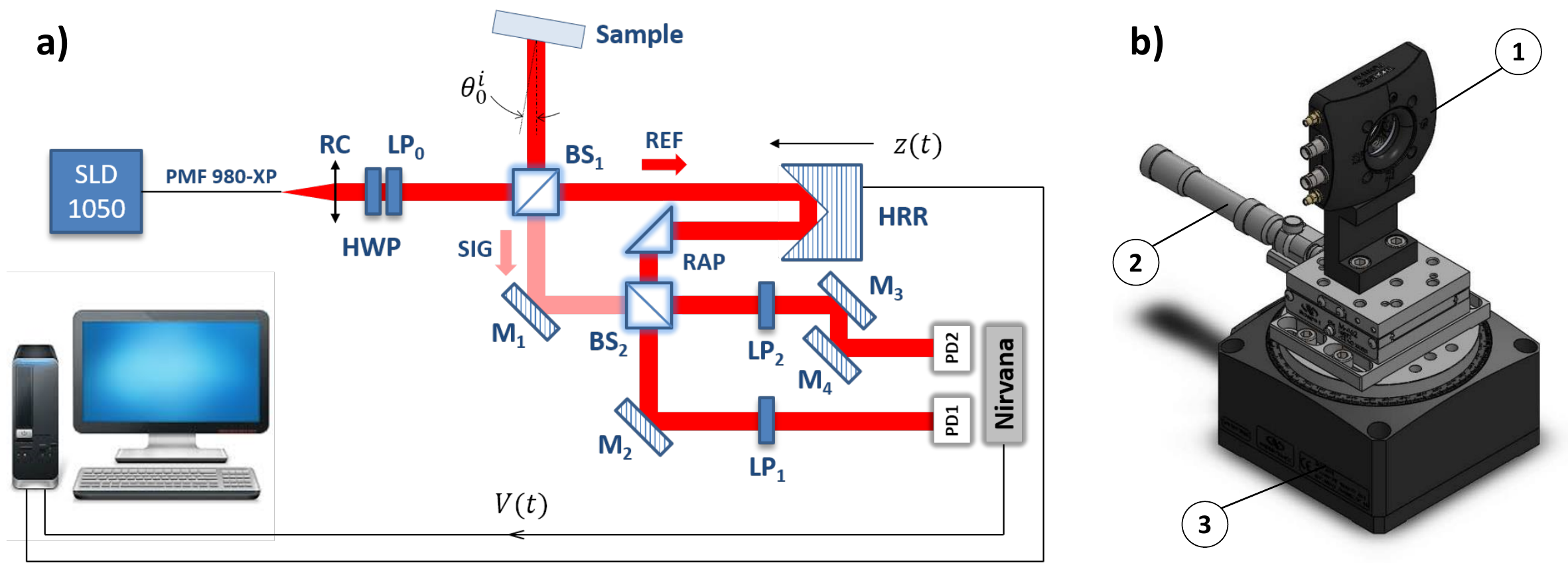}
\caption{\label{fig:Bench_Scheme}a) Schematic diagram of the BARRITON set-up - b) 3D view of the sample holder [1 - Piezoelectric gimbal mount, 2 - Manual horizontal translation stage, 3 - Motorized rotation stage].}
\end{figure*}

A schematic of our experimental setup, named BARRITON (for Back-scattering And Retro-Reflection by InterferomeTry with lOw cohereNce) is given in Fig.~\ref{fig:Bench_Scheme}a.

The linearly polarized divergent light beam provided by a fibered superluminescent diode (SLD) centered at 1050 nm (light power about 70 mW, spectral bandwidth about 80 nm for a maximum driving current of 1000 mA) is collimated by an off-axis parabolic mirror of 7 mm focal length. The orientation of the polarization direction of this weakly divergent Gaussian light beam ($w_f=0.67$ mm, $\theta_f\sim0.03$ degrees) is finely tuned with respect to the vertical transmission axis of a LP$_0$ nanoparticle polarizer using a rotating half-wave plate (HWP).
This polarized beam is divided into two perpendicular sub-beams by the non-polarizing cube splitter BS$_1$: the reflected sub-beam is directed towards the sample and the light flux retro-reflected or back-scattered by the sample is transmitted by the same cube splitter BS$_1$ (SIG arm). The transmitted sub-beam (REF arm) is simultaneously retro-reflected and laterally shifted by a hollow retro-reflector HRR mounted on a motorized translation stage whose axis of movement is coincident with the direction of the REF beam (delay line). This beam then passes through a right angle prism (RAP) and is combined with the SIG beam through the non-polarizing cube splitter BS$_2$. The two output channels of this cube splitter are finally detected by the two photodiodes (PD1 and PD2) of a balanced receiver (Nirvana)\cite{Hobbs_1997}.

As shown in Fig.~\ref{fig:Bench_Scheme}b, the sample is placed in a piezoelectric gimbal mount with two axes that each rotate its front side around a gimbal axis, holding the center of the sample in a fixed position. Each axis is capable of scanning an angular range of 30 mrad with a resolution of 0.05 $\mu$rad. This system allows the front side of the sample to be positioned perpendicular to the incident beam. The variation of the angle of incidence (AOI) of the light beam on the sample is ensured by a motorized rotation stage with vertical axis (Newport RGV100BL-S), characterized by a minimum incremental motion of 0.1 mdeg. A manual horizontal translation stage, located between the gimbal mount and the rotation stage, allows to adjust the axial position of the front face of the sample so that it contains the rotation axis of this motorized stage.

The balanced receiver delivers a voltage $V$ proportional to the difference between the electrical currents $I_1$ and $I_2$ of the two photodiodes, and which corresponds to the following expression\cite{Khan_2021,Lequime_2022}
\begin{equation}
V=G\mathcal{T}\thinspace\Re\left\{\int\limits_{0}^{\infty}S(f)\mathcal{P}(f)\rho(f)\thinspace e^{-ik\Delta L}\thinspace df\right\}
\end{equation}
where $f$ is the frequency of the optical field, $S(f)$ is the spectral dependence of the photodiode responsivity, $\mathcal{P}(f)$ is the power spectral density of the light source, $\rho(f)$ is the coherent coefficient of retro-reflection/backscattering of the sample\cite{Amra_2021,Khan_2021}, $k$ is the wave vector, $\Delta L$ is the optical path difference between SIG and REF channels, $G$ is the transimpedance gain of the balanced receiver, and $\mathcal{T}$ is a global transmission factor characteristic of the set-up\cite{Khan_2021}.

\begin{figure*}
\includegraphics[width=0.85\textwidth]{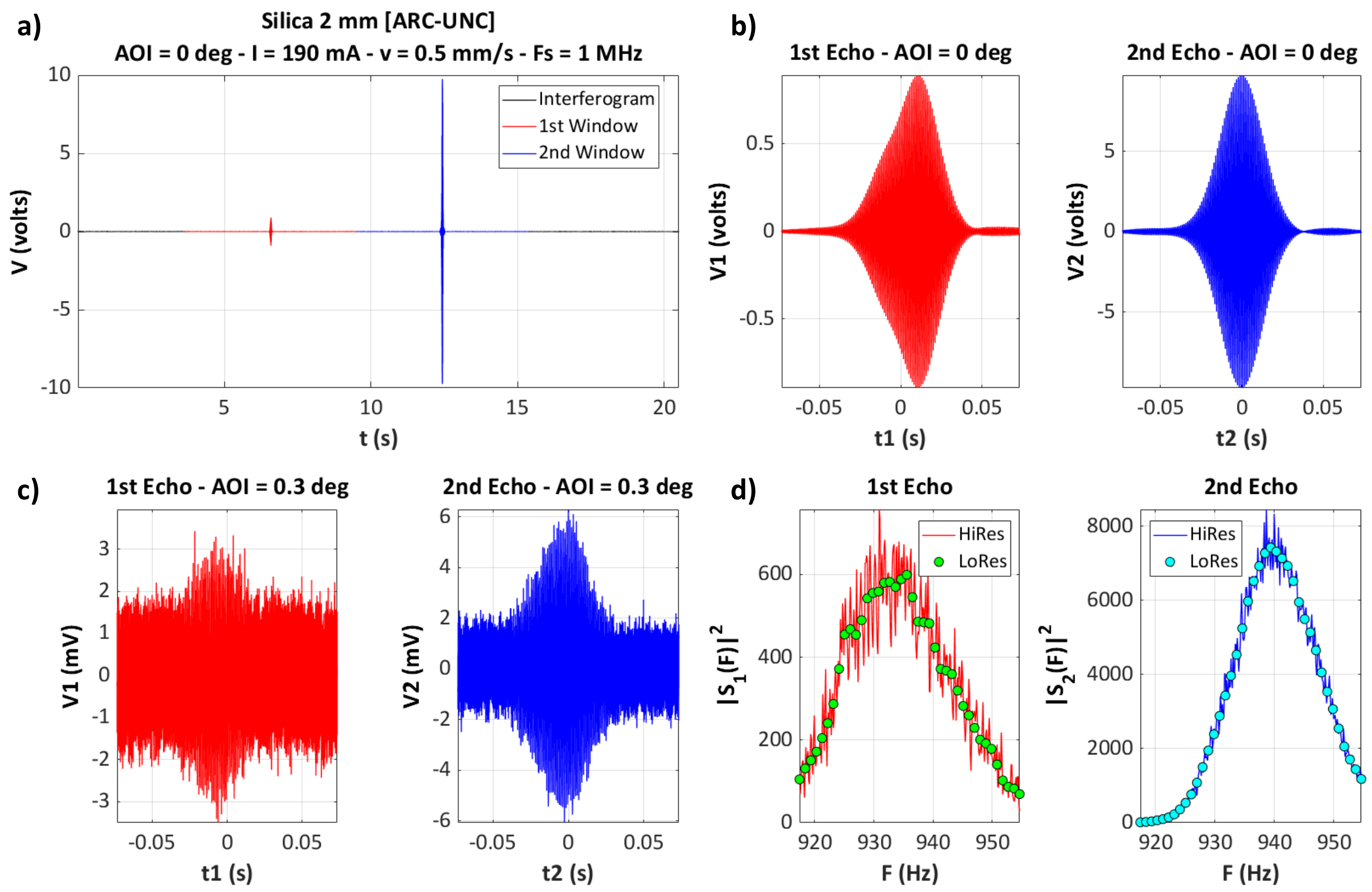}
\caption{\label{fig:Interferogram} Examples of interferogramms recorded with BARRITON when the front face of the sample corresponds to the antireflective coated face of a 2 mm thick silica wafer (translation speed $v$ = 0.5 mm/s) - a) Full interferogram recorded at AOI 0 degree (digitizing range 10 Volts) - b) Magnified view of the two echoes recorded at AOI 0 degree (coated face in red, uncoated face in blue, digitizing range 10 Volts) - c) Magnified view of the two echoes recorded at AOI 0.3 degrees (coated face in red, uncoated face in blue, digitizing range 0.1 Volts) - d) Square modulus of the DFTs of the interferogram recorded at AOI 0.3 degrees (High Resolution and Low Resolution modes).}
\end{figure*}

This voltage $V$ is digitized over 16 bits with a sampling rate $F_s$ of 1 MHz, while the hollow retro-reflector HRR is translated at a constant speed $v$ along the $z$ axis [$\Delta L=2vt$]. Figure \ref{fig:Interferogram}a shows the time dependence of this digitized voltage when the front face of the sample corresponds to the coated face of a 2 mm thick silica wafer (0 degree AOI, $v$ = 0.5 mm/s). Figure \ref{fig:Interferogram}b (respectively \ref{fig:Interferogram}c) shows a magnified view of the echoes associated with each of the faces of this silica wafer (in red, the coated front face; in blue, the uncoated back face), in retroreflection for an AOI of 0 degree (respectively in backscattering, for an AOI of 0.3 degrees). The digitizing range (10 Volts, 1 Volt, 0.1 Volt) is automatically adapted to the amplitude of the strongest echo.

The two graphs of Figs. \ref{fig:Interferogram}b and \ref{fig:Interferogram}c show that the signal-to-noise ratio of the interferogram decreases very rapidly with the tilt angle of the sample, because the level of the backscattered flux is much lower than that of the retroreflected flux. Moreover, the profile of each of these two echoes does not give any information about the spectral dependence of the light back-scattered by each interface of the sample. To overcome these two limitations, we use a data processing transposed from the one implemented in Fourier transform spectrometry\cite{Lequime_2022}.

We first isolate in the interferogram recorded at $\theta_0^{i}$ two temporal windows centered on each of the two echoes and with a full width $\Delta T$. Then we perform a discrete Fourier transform (DFT) of the corresponding data set
\begin{equation}
S_m(F_l,\theta_0^{i})=\sum\limits_{k=-N/2}^{k=N/2-1}V_m(t_k,\theta_0^{i})\thinspace e^{-2i\pi F_lt_k}\thinspace dt
\label{eq:TFVm}
\end{equation}
where
\begin{equation}
dt=\frac{1}{F_s}=\frac{\Delta T}{N-1}\quad\text{;}\quad t_k=t_m+k.dt
\end{equation}
and
\begin{equation}
F_l=\frac{2v}{c}f_l=l.dF\quad\text{;}\quad dF=\frac{1}{(N-1)dt}=\frac{1}{\Delta T}
\end{equation}
If the width $\Delta T$ of the time window is less than or equal to the time interval $\Delta T_e$ between the two echoes, it can be shown\cite{Lequime_2022} that the DFT associated with the echo $m$ is proportional to the complex conjugate of the corresponding retro-propagation coefficient $\rho_m$, i.e.
\begin{equation}
S_m(F_l,\theta_0^{i})=G\mathcal{T}\frac{c}{4v}S(f_l)\mathcal{P}(f_l)\rho_m^*(f_l,\theta_0^{i})\quad m=1,2
\end{equation}
Figure \ref{fig:Interferogram}d shows the frequency dependence of $|S_m(F_l)|^2$ for the two echoes recorded at 0.3 degrees when the DFT is calculated with two different processing modalities: HiRes ($\Delta T=\Delta T_e$) and LoRes ($\Delta T=\Delta T_e/10$). Note the significant improvement of the signal-to-noise ratio resulting from the calculation of these DFTs (the white noise of quantum origin which affects the interferograms is dispersed over the whole spectrum and its importance is therefore automatically reduced). It is also important to underline the additional gain brought by the use of LoRes mode, but equally its negative impact on the spectral resolution. This one is indeed inversely proportional to $\Delta T$ and equal to 0.2 nm in HiRes mode and 2 nm in LoRes mode.

For example, let us suppose that the front face of the sample is the uncoated one. Under these conditions, we have\cite{Lequime_2022}
\begin{equation}
R_2(f_l)=\frac{R_1(f_l)}{[1-R_1(f_l)]^2}\frac{|S_2(f_l,0)|^2}{|S_1(f_l,0)|^2}
\label{eq:R2}
\end{equation}
\begin{equation}
|\rho_m(f_l,\theta_0^{i})|^2=R_m(f_l)\frac{|S_m(f_l,\theta_0^{i})|^2}{|S_m(f_l,0)|^2}\quad m=1,2
\end{equation}
where we assumed that the angle dependence of the transmission coefficient of the front face is negligible. Similar relationships can be established for the case where the front side of the sample is the coated side.

The sample we used for this experimental demonstration is a high quality component purchased from \textsc{Light Machinery} \cite{LightMachinery_2022} and consisting of a 2 mm thick, 25 mm diameter, 7980 A grade fused silica window with a wedge angle of 0.6 arc seconds and one side coated with a V-shaped antireflective coating centered at 1055 nm.

\begin{figure*}
\includegraphics[width=1.0\textwidth]{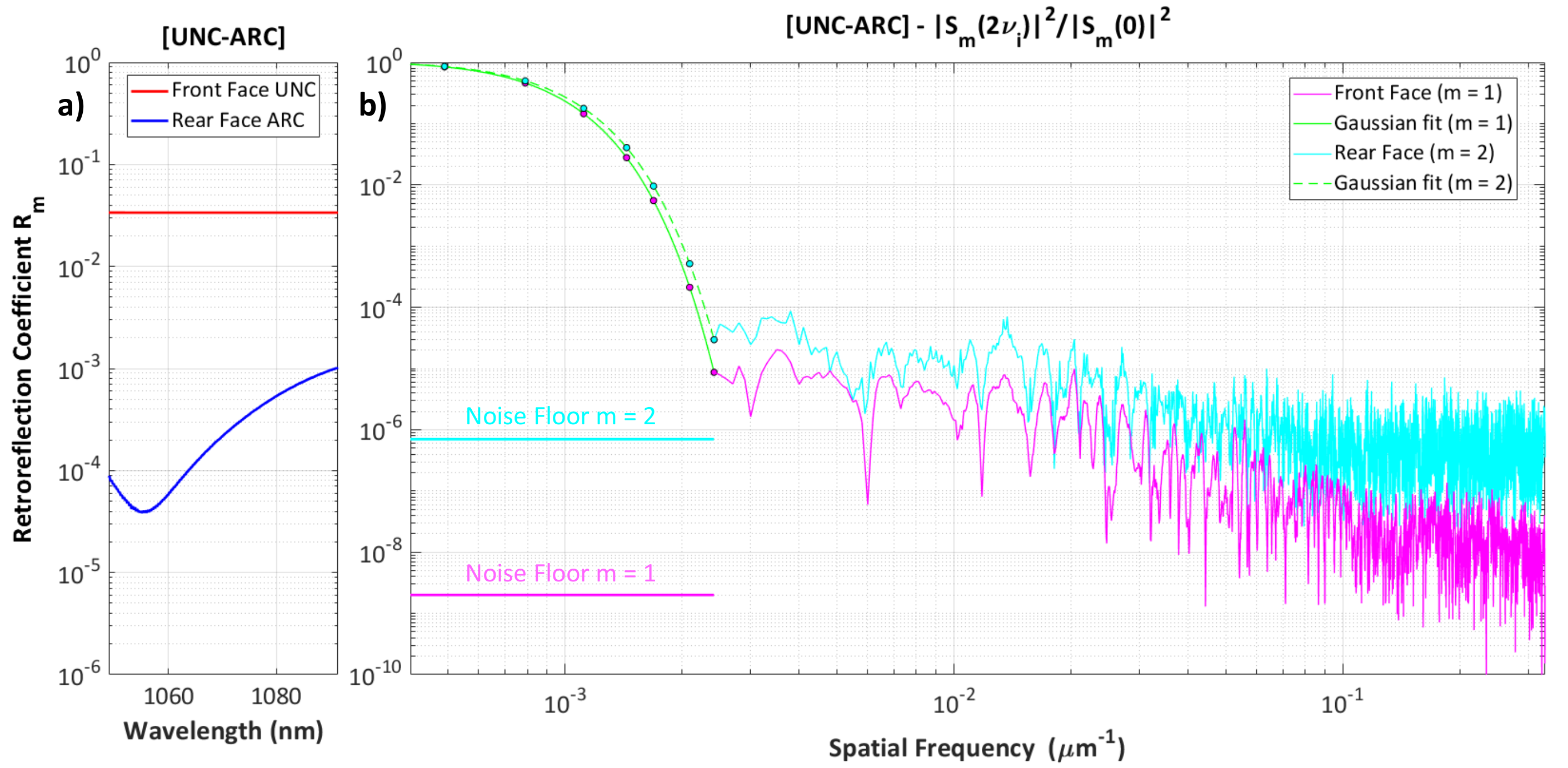}
\caption{\label{fig:Scattering_UNC_ARC}Results of the experimental measurement of the retroreflection and backscattering properties of the silica wafer sample (front face uncoated, rear face coated) - a) Spectral dependence of the retroreflection coefficient of each of the two faces - b) Evolution of the ratios $|S_m(\nu)|^2/|S_m(0)|^2$ ($m=1,2$) as a function of the spatial frequency $\nu=2\nu_i$ (front face, magenta dots and magenta solid line; rear face, cyan dots and cyan solid line).}
\end{figure*}

The measurements are performed with the following operating conditions : SLD driving current $I$ of 190 mA, translation speed $v$ of 0.5 mm/s, sampling frequency $F_s$ of 1 MHz, angle pitch of 10 mdeg, and angular range from 0 to 10 degrees. Consider for example the [UNC-ARC] configuration, in which the front face of the sample is not coated. The coefficient $R_1$ has then the expression 
\begin{equation}
R_1(f_l)=\left[\frac{1-n_s(f_l)}{1+n_s(f_l)}\right]^2
\end{equation}
where $n_s(f_l)$ is the refractive index of the silica substrate. The spectral dependence of this refractive index is perfectly known, which allows the value of $R_1(f_l)$ to be calculated at any frequency $f_l$. The use of equation (\ref{eq:R2}) allows us to deduce $R_2(f_l)$ using the results of our measurements at zero incidence (see Fig.~\ref{fig:Scattering_UNC_ARC}a).

In the case of the quantities $|\rho_m(f_l,\theta_0^{i})|^2$ which describe the spectral and angular dependence of the light flux backscattered by the $m$th interface, the presentation of our experimental results requires beforehand the statement of some theoretical considerations. The elementary light fields scattered by each interface $m$ towards the incident medium may be described by a general expression of the type\cite{Amra_2021}
\begin{equation}
\vec{\mathbb{A}}_m^{-}(\vec{\nu}_d)=\sum\limits_{j=0}^{p_m}C_{m,j}^{-}[\hat{h}_{m}\star\hat{s}_e]_{\vec{\nu}_d,\vec{\nu}_i}\vec{\mathbb{A}}_i
\end{equation}
where $p_m$ is the number of layers deposited on the interface $m$, $C_{m,j}^{-}$ are coefficients related to the optical structure of the component, $\hat{h}_m$ is the Fourier transform of the $m$ interface height profile (for simplicity, we assumed that the profiles of the layers are identical to the one of the interface on which they are deposited), $\hat{s}_e$ is the Fourier transform of the spatial distribution of the incident light field at the surface of the component, and $\vec{\nu}_d$ (respectively $\vec{\nu}_i$) is the spatial frequency of the scattered (respectively incident) field.

For the backscattered field, we have
\begin{equation}
\vec{\nu}_d=-\vec{\nu}_i\enskip\text{with}\enskip\vec{\nu}_i=(\nu_i,0)\enskip\text{and}\enskip\nu_i=\frac{\sin\theta_0^{i}}{\lambda}
\end{equation}
For a weakly divergent Gaussian beam, $\hat{s}_e$ can be assimilated to a Dirac function $\delta(\vec{\nu}_d-\vec{\nu}_i)$, and the amplitude of the backscattered fields satisfies the following equation
\begin{equation}
\vec{\mathbb{A}}_m^{-}(-\vec{\nu}_i)\propto\hat{h}_m(\vec{\nu}_d-\vec{\nu}_i)=\hat{h}_m(-2\vec{\nu}_i)=\hat{h}_m^*(2\vec{\nu}_i)
\end{equation}
where the last equality is due to the Hermitian symmetry of $\hat{h}_m$ ($h_m$ is real). This means that the results of our measurements should be plotted not as a function of the angle of incidence $\theta_0^{i}$ or wavelength $\lambda$, but as a function of twice the spatial frequency $\nu_i$ associated with the incident wave. This is particularly important because it shows that the number of independent data we have for the measurement of the spatial frequency dependence of the quantities $\rho_m$ is in fact equal to the product of the number of angular values (here, 1000) by the number of spectral values (in our case, about 300 in HiRes mode and 30 in LoRes mode), which makes it possible to improve both the signal-to-noise ratio of the measurement and its resolution in spatial frequency.

Figure \ref{fig:Scattering_UNC_ARC}b shows the results obtained when the step $\delta\nu$ is equal to $10^{-4}$ $\mu\text{m}^{-1}$. On these curves, we see appear, beyond the specular response ($\nu\geqslant 2.5\times 10^{-3}$ $\mu\text{m}^{-1}$), a speckle phenomenon whose grain size is governed by both the divergence of the incident beam and the spectral resolution of the measurement.

In the spatial frequency domain corresponding to retroreflection ($0\leqslant\nu\leqslant 2.5\times 10^{-3}$ $\mu\text{m}^{-1}$), the analytical calculation can be completed, and shows that the spatial frequency dependence of the ratio $|\rho_m(\nu,f_l)|^2/R_m(f_l)$ is described by\cite{Lequime_2019}
\begin{equation}
\frac{|\rho_m(\nu,f_l)|^2}{R_m(f_l)}=e^{-\pi^2 w_m^2\nu^2}\enskip\text{where}\enskip w_m=\frac{w_s}{\sqrt{1+\cos^2 2\theta_0^{i}}}
\label{eq:GaussianFit}
\end{equation}
The two green curves in Fig.~\ref{fig:Scattering_UNC_ARC}b correspond to the best fits of the experimental data based on the relation (\ref{eq:GaussianFit}) and allow to determine the values of $w_m$, i.e. $w_1=496$ $\mu$m and $w_2=469$ $\mu$m. We can then deduce the value of $w_s$ which corresponds to the half-size of the beam at the sample. The value obtained, $w_s=682\pm50$ $\mu$m, is in very good agreement with the one ($w_s=706$ $\mu$m) resulting from the application of the formulas of propagation of the Gaussian beams between the exit of the reflective collimator and the sample, that is
\begin{equation}
w_s=\sqrt{1+\left(\frac{d_s}{z_R}\right)^2}\enskip\text{with}\enskip z_R=\frac{\pi w_f^2}{\lambda}\text{ and }w_f=\frac{f\lambda}{\pi w_0}
\end{equation}
where $w_0=3.5$ $\mu$m @ $\lambda=1070$ nm, $f=7$ mm, and $d_s=375$ mm.

The knowledge of this quantity will allow us to transform the curves represented on the graph of Fig.~\ref{fig:Scattering_UNC_ARC}b into BRDF $\cos\theta$. Indeed, we have\cite{Khan_2021,Was_2022}
\begin{equation}
\text{BRDF}_m\thinspace\cos\theta=\text{ARS}_m(\lambda,\nu)=\frac{\pi w_s^2}{\lambda^2}R_m(\lambda)\frac{|S_m(\nu)|^2}{|S_m(0)|^2}
\end{equation}
For the uncoated face ($m=1$), all measurements are made with a satisfactory signal-to-noise ratio, demonstrating that our set-up is capable of measuring ARS values as low as $10^{-4}\text{ sr}^{-1}$ over a spatial frequency range from 0 to $3.5\times 10^{-1}$ $\mu\text{m}^{-1}$. For the coated face ($m=2$), the highest spatial frequency is reduced to $5\times 10^{-2}$ $\mu\text{m}^{-1}$, due to the presence of the antireflective coating.

A theoretical analysis of the different noise sources\cite{Lequime_2022} shows that it is possible to significantly improve the detection sensitivity of this set-up, by optimizing different parameters, such as the translation speed $v$, the bandwidth $B$ of the balanced receiver or the optical power corresponding to the saturation of the photodiodes. A gain of 500 seems quite accessible, which would allow the detection of ARS as low as $2\times 10^{-7}$ $\text{sr}^{-1}$.

With such detection sensitivity, it would also become possible to accurately measure the spectral dependence of the phase of the field backscattered from the first interface, using the following relation
\begin{equation}
\text{arg}[\rho_1(f_l,\theta_0^{i})]=-\text{arg}[S_1(f_l,\theta_0^{i})]
\end{equation}
which paves the way for a reconstruction of the effective height profile of the optical interface under observation.

\section*{Acknowledgments}
This work is part of the StrayLight Working Group for the Laser Instrument Group of the LISA Consortium. The authors would like to thank the AMIdex Talents Management Program, Aix
Marseille University and the French National Space Center (CNES) for their financial support, and the Virgo Collaboration for fruitfull discussions.
\section*{Disclosures}
The authors declare no conflicts of interest.
\section*{Data availability}
Data underlying the results presented in this paper are not publicly available at this time but may be obtained from the authors upon reasonable request.

\bibliographystyle{aipnum4-1.bst}
\section*{References}
\bibliography{Barriton_APL_V2}

\end{document}